\documentclass[lettersize,journal]{IEEEtran}
\usepackage{amsmath,amsfonts}
\usepackage{algorithmic}
\usepackage{array}
\usepackage[caption=false,font=normalsize,labelfont=sf,textfont=sf]{subfig}
\usepackage{textcomp}
\usepackage{stfloats}
\usepackage{hyperref}
\usepackage{url}
\usepackage{verbatim}
\usepackage{graphicx}
\usepackage{multirow}
\usepackage{tipa}
\usepackage{xspace}
\def\BibTeX{{\rm B\kern-.05em{\sc i\kern-.025em b}\kern-.08em
    T\kern-.1667em\lower.7ex\hbox{E}\kern-.125emX}}
\usepackage{balance}
\begin{document}

\title{Coding Speech through Vocal Tract Kinematics}
\author{Cheol Jun Cho, Peter Wu, Tejas S. Prabhune, Dhruv Agarwal, and Gopala K. Anumanchipalli
\thanks{All authors are with the Department of Electrical Engineering and Computer Sciences at University of California, Berkeley (e-mail: cheoljun@berkeley.edu; peterw1@berkeley.edu; prabhune@berkeley.edu; dhruvagarwal@berkeley.edu; gopala@berkeley.edu).\\Corresponding author: Cheol Jun Cho; Gopala K. Anumanchipalli}}

\markboth{}
{How to Use the IEEEtran \LaTeX \ Templates}

\maketitle

\begin{abstract}
Vocal tract articulation is a natural, grounded control space of speech production. The spatiotemporal coordination of articulators combined with the vocal source shapes intelligible speech sounds to enable effective spoken communication. Based on this physiological grounding of speech, we propose a new framework of neural encoding-decoding of speech -- Speech Articulatory Coding (SPARC). SPARC comprises an articulatory analysis model that infers articulatory features from speech audio, and an articulatory synthesis model that synthesizes speech audio from articulatory features. The articulatory features are kinematic traces of vocal tract articulators and source features, which are intuitively interpretable and controllable, being the actual physical interface of speech production. An additional speaker identity encoder is jointly trained with the articulatory synthesizer to inform the voice texture of individual speakers. By training on large-scale speech data, we achieve a fully intelligible, high-quality articulatory synthesizer that generalizes to unseen speakers. Furthermore, the speaker embedding is effectively disentangled from articulations, which enables accent-perserving zero-shot voice conversion. To the best of our knowledge, this is the first demonstration of universal, high-performance articulatory inference and synthesis, suggesting the proposed framework as a powerful coding system of speech.
\end{abstract}

\begin{IEEEkeywords}
speech coding, speech synthesis, articulatory synthesis, speech inversion, acoustic-to-articulatory inversion, electromagnetic articulography.
\end{IEEEkeywords}

\section{Introduction}
Humans naturally produce intelligible speech by controlling articulators on the vocal tract. Such vocal tract articulation has long been claimed to be the physiological ground of speech production in various aspects \cite{browman1992articulatory}. The source-filter theory of speech describes articulation as shaping the vocal cavity to implement filters on source, or glottal flow, to create speech sounds \cite{chiba1958vowel, fant1971acoustic}. Articulatory phonetics and phonology have explained the basis of speech in terms of the coordination of articulators, identifying some canonical articulators that can determine the phonetic properties \cite{maeda1990compensatory, browman1992articulatory, international1999handbook}. In cognitive neuroscience, the speech sensorimotor cortex has been proven to represent continuous, real-time vocal tract articulation while naturally speaking, suggesting the vocal tract articulation as a cognitive basis of speech production \cite{chartier2018encoding, anumanchipalli2019speech, cho2023neural}. 

Furthermore, the recent findings by Cho et al. \cite{cho2023evidence,cho2023self} suggest that articulatory inversion naturally emerges from self-supervised learning (SSL) of speech. When probed on articulatory kinematics measured by electromagnetic articulography (EMA), the feature representation of the recent speech SSL models (e.g., HuBERT \cite{hsu2021hubert}) is highly correlated with EMA, where high-fidelity articulation can be reconstructed by a simple linear mapping from speech SSL features \cite{cho2023evidence}. This suggests that the articulatory inference is a natural solution of SSL of speech for abstracting speech information. This emergent property is further shown to be universal across speakers, dialects, and even languages \cite{cho2023self}. Together, these suggest that the biophysical, articulatory representation of speech is a shared coding principle in both biological and artificial intelligence of speech. This convergence raises an interesting question -- \textit{Can we represent any arbitrary speech using articulatory features?}

\begin{figure}[!t]
\centering
\includegraphics[width=.95\linewidth]{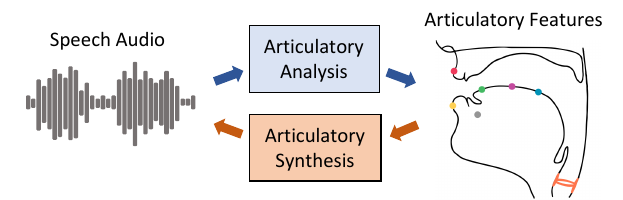}
\vspace{-5pt}
\caption{The Speech Articulatory Coding (SPARC) framework. It encodes speech using articulatory features (analysis) and decodes these features back into speech (synthesis). 
}
\label{fig:general}
\vspace{-10pt}
\end{figure}

Previous studies have demonstrated that intelligible speech can be synthesized from articulatory features \cite{birkholz2013prev_artsynth, krug2021vocaltractlab, wu2022artsynth, kim2023style, gao2024copysynthesis}, and combined with acoustic-to-articulatory inversion (AAI), resynthesis frameworks have shown the potential of articulatory features as viable intermediate for speech coding systems \cite{wu2023speakerind, gao2024copysynthesis}. However, these previous methods are limited to a fixed set of speakers and the quality is still far behind the commercial speech synthesis models. This absence of a universal, generalizable framework has significantly limited the practical utility of articulatory-based speech coding. 

Here, we first demonstrate a high-performance, universal articulatory encoder and decoder that can scale and generalize across an indefinite number of speakers. We leverage the universal articulatory inference by speech SSL \cite{cho2023self} to build a generalizable articulatory encoder that transforms speech into vocal tract kinematic traces in a template articulatory space. The template articulatory space is agnostic to individual anatomical differences which are compensated by a separate speaker identity encoder. By training a synthesis model, or a vocoder, on a large-scale dataset, we achieve a universal articulatory vocoder that can generate fully intelligible, high-quality speech from any speaker's articulation. Furthermore, the speaker identity encoder successfully disentangles speaker identity from articulation, which is demonstrated by a zero-shot, accent-preserving voice conversion. By closing the loop of articulatory encoding and decoding, we propose a novel, speech science guided encoding-decoding framework of speech 
-- Speech Articulatory Coding (SPARC) (Fig. \ref{fig:general}).\footnote{The code is available at \url{ https://github.com/Berkeley-Speech-Group/Speech-Articulatory-Coding}.} The SPARC framework shows a minimal loss of intelligibility and quality compared to the original speech audio.\footnote{Audio samples: \url{https://berkeley-speech-group.github.io/sparc-demo}}

Compared to existing neural coding of speech \cite{choi2021nansy, choi2022nansy++, zeghidour2021soundstream, defossez2022encodec, ju2024naturalspeech}, representing speech as articulatory features has following benefits:

\begin{itemize}
\item{\textbf{Low-dimensionality}: The articulatory features have only 14 channels with 50 Hz sampling rate. 
}
\item{\textbf{Interpretability}: Each channel corresponds to the actual physical articulator on the vocal tract, which can be intuitively interpretable by visualization.}
\item{\textbf{Controllability}: The features can be naturally controlled by the same principle as speech production.}
\item{\textbf{Universality}: The articulatory encoding is universal across speakers and disentangled from individual anatomical variance.}
\end{itemize}

With these unique benefits, we demonstrates empirical evidence of the promising potential of SPARC as a valid, novel coding framework for speech.
\section{Related Work}
\subsection{Electromagnetic Articulography}


Electromagnetic articulography (EMA) measures time-varying displacements of vocal tract articulators synchronously while speaking. Typically, sensors are placed on the upper lip (UL), lower lip (LL), lower incisor (LI), tongue tip (TT), tongue blade (TB), and tongue dorsum (TD) (Fig. \ref{fig:pipeline}) \cite{rebernik2021review}. A combination of displacements of these articulators on the midsagittal plane configures a place of articulation, and combined with source information, or manner of articulation, it shapes a phonetic content of speech. As the traces are continuously collected in real-time, the EMA data naturally reflect phoneme contextualization (coarticulation) and individual tendencies in pronunciations (accents). Given these properties, EMA has been widely accepted for studying articulatory bases of speech, providing biophysical evidence for many linguistic or cognitive theories of speech production \cite{ browman1992articulatory, rebernik2021review, chartier2018encoding}. However, EMA has been significantly limited to scale due to the complicated nature and high cost of the collection procedure. 

\subsection{Acoustic-to-Articulatory Inversion} \label{relatedwork:AAI} To replace the complicated data collection procedure, acoustic-to-articulatory inversion (AAI) models have been actively developed to predict EMA directly from speech audio \cite{ghosh2010generalized,  ghosh2011subject, liu2015deep, chartier2018encoding, anumanchipalli2019speech, wu2023speakerind, attia2023improving, siriwardena2023secret, gao2024copysynthesis}. However, the individual variance in vocal tract anatomy across speakers induces inconsistent placements of sensors, which has posed a significant barrier to developing a model that can generalize to unseen speakers \cite{rahim1993annartsynth, wu2023speakerind, attia2023improving, siriwardena2023secret}. Despite such variability, a canonical basis of articulation is suggested to exist, which is agnostic to individual vocal tract anatomy \cite{browman1992articulatory, seneviratne2019multi, wu2023speakerind, cho2023self}. In fact, Cho et al. \cite{cho2023self} demonstrated that a linear affine transformation can geometrically align one speaker's articulatory system to another's. This suggests that individual articulatory spaces are lying on the same linear space so that an articulatory space of any speaker can be a hypothetical universal template space of articulation. We empirically validate this statement by using a single-speaker AAI model as a universal articulatory encoder for our coding framework. 

\subsection{Articulatory Synthesis} 

Articulatory synthesis aims to generate speech audio from articulatory features. A century of efforts have been made to build articulatory synthesizers for basic research of speech \cite{dudley1939synthetic, dudley1939remaking, dunn1950calculation, stevens1953electrical, rosen1958dynamic, mermelstein1973articulatory, rubin1981articulatory, maeda1982digital, scully1990articulatory}. Several methods have been proposed for improving intelligibility and quality
, demonstrating broader use cases including text-to-speech (TTS) \cite{krug2021vocaltractlab}, prosody manipulation \cite{aryal2014accent, birkholz2017manipulation}, speech denoising \cite{georges2021artpriorvae}, and speech brain-computer interfaces (BCIs) \cite{anumanchipalli2019speech}. Some of these works utilize deep learning models to map articulatory features to acoustic features, which are then converted to audio using pretrained acoustic synthesizers \cite{aryal2016data, anumanchipalli2019speech, georges2020artsynthlpcnet, kim2023style}. A recent study shows that a GAN-based generative model can directly synthesize speech waveform from articulatory features with high intelligibility \cite{wu2022artsynth}. However, to our knowledge, none of the existing approaches has achieved industrial-level performance, which requires high intelligibility, quality, and generalizability across unseen speakers.

\subsection{Neural Coding of Speech} 

Many deep learning methods have been proposed to learn data-driven representations of speech. Various autoencoder-based frameworks have been suggested to jointly train encoders that compress audio into low-bitrate discrete units \cite{van2017neural, zeghidour2021soundstream, defossez2022encodec} or decompose speech into different factors \cite{choi2022nansy++, zhang2023speechtokenizer, ju2024naturalspeech}, and decoders that reconstruct speech from encoded features with minimal loss of information. Also, pretrained speech SSL models have been utilized to extract rich linguistic content of speech, and synthesizers are trained to restore speech audio from those features \cite{choi2021nansy, polyak2021resynthesis, choi2022nansy++,  lee2022hierspeech, lee2023hierspeech++, guo2023quickvc}. These SSL-based methods often utilize separate source modeling (e.g., pitch) and speaker encoding, since SSL model encoders tend to marginalize out acoustic and speaker information \cite{chen2022spkinfo, pasad2023comparative}. We categorize all these kinds of closed-loop frameworks utilizing neural networks for both encoding and decoding as \emph{neural coding of speech}. Though the existing methods achieve high fidelity in representing speech audio, the intermediate speech codes significantly lack interpretability, and embeddings of those codes are often high-dimensional. 

Here, we propose articulatory features as interpretable, controllable, and grounded coding of speech that can fully represent any arbitrary speech. A similar concurrent work is proposed towards interpretable speech representations, which utilizes sparse phonetic posteriorgrams instead of articulatory features \cite{morrison2024sppg}. 

\begin{figure*}[t]
\centering
\includegraphics[width=0.9\linewidth]{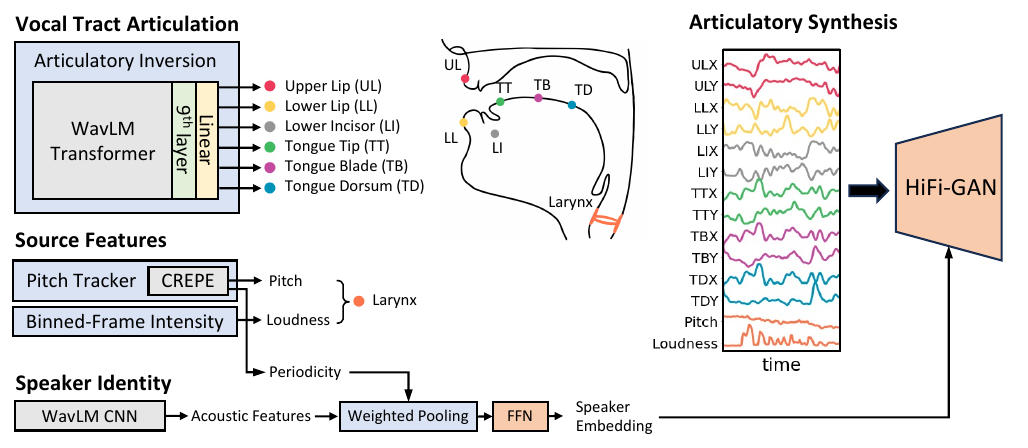}
\vspace{0pt}
\caption{Pipeline of articulatory analysis and synthesis. The articulatory analysis is composed of vocal tract articulation, source features, and speaker embedding, which are then fed to the synthesizer (HiFi-GAN) in the synthesis pipeline. The modules colored with orange (FFN and HiFi-GAN) are updated while training the synthesis model and other modules are fixed. 
}
\label{fig:pipeline}
\end{figure*}

\section{Methods}

To bridge the interpretability and controllability gap in current neural speech coding systems, we propose neural articulatory inversion and synthesis as a new type of coding system that can provide an interpretable and controllable coding of speech.

\subsection{Articulatory Analysis}

In SPARC, speech is encoded as factors obtained by an analytic framework that infers three different components of speech: vocal tract articulation, source features, and speaker identity (Fig. \ref{fig:pipeline}). The first two are based on pre-trained analysis models that provide kinematic traces of the physical articulators, and vocal source features. The last component, speaker identity, is inferred by a model jointly trained with the synthesizer.

\subsubsection{Vocal Tract Articulation}
\label{methods:inversion} 
 Based on the findings by \cite{cho2023self} elaborated in \S \ref{relatedwork:AAI}, we propose to use a single speaker's EMA as a template articulatory space to represent speaker-generic articulatory kinematics. We selected one of the largest single-speaker EMA datasets, MNGU0 \cite{mngu}, that includes 75 minutes of EMA collected while reading newspapers aloud. This dataset is widely accepted and verified in many studies, given a fine signal quality carefully controlled by the authors. We claim all speakers' articulations can be represented on this single-speaker EMA space without losing information that contributes to the intelligibility of speech. That is, EMA represents phonetic content in a way that can be detached from the variance of vocal tract anatomical structure across individuals. 

We use the SSL-linear AAI approach proposed by \cite{cho2023evidence, cho2023self}. The SSL-linear model is built by training a linear mapping from SSL features to EMA, while keeping the SSL encoder weights frozen. This simple mapping can effectively find a linear subspace in the SSL feature space which is highly correlated with EMA, as shown by previous probing studies \cite{cho2023evidence,cho2023self}. We use the WavLM Large model \cite{chen2022wavlm}, which shows the highest correlation amongst speech SSL models as reported in \cite{cho2023evidence}. Note that the linear head is the only fitted part here, thus, maintaining the generalization capacity of the WavLM encoder that is attained by pretraining on large-scale speech data and adversarial data augmentation \cite{chen2022wavlm}. Furthermore, the speaker information tends to diminish after a few early layers \cite{chen2022spkinfo}, which indicates that the mapping can be speaker-agnostic, further contributing to multi-speaker generalizability. 

The original 200 Hz EMA data is downsampled to 50 Hz to match the sampling rate of the SSL features, and each channel is z-scored within utterances. The 9th layer of the WavLM Transformer encoder is used to extract speech features for the inversion, where the input audio has 16000 Hz sampling frequency, zero mean, and unit variance. A low-pass filter is applied to the features to remove high-frequency noise using the Butterworth fitler with order of 5, where the frequency threshold is set as 10 Hz. The linear inversion model is trained by ordinary least squares. All data in MNGU0 dataset are used for training the main model after selecting the best layer using cross-validation. See Appendix A.1 for the details of the layer selection procedure. The resulting AAI model outputs 12 channels of EMA (X and Y axis of each of 6 articulators) with 50 Hz frequency.

We claim the proposed AAI model can be universal based on two hypotheses. First of all, all individual articulatory spaces lie on the same linear space. Articulatory space is defined as a vector space where each basis corresponds to physical location of a specific articulator. The space is speaker dependent as individual has different vocal tract anatomy. However, an affine transformation between two spaces exists that spatially aligns different vocal tract structures. As elaborated in \S \ref{relatedwork:AAI}, empirical evidence is demonstrated by \cite{cho2023self} and we also report in \S \ref{sec/univ_aai}. This suggests that despite variable vocal tract anatomies there exists a common basis to which all individual articulations can be registered. All we need is a high quality measurement of any articulatory space, which we claim the MNGU0 is sufficient for this purpose. 

The second hypothesis is that the SSL-linear AAI is speaker agnostic. This again requires the following propositions to hold: there exists a subspace in SSL features that is agnostic to voice identity, and a linear mapping used to project the features to the subspace is also speaker independent. This means that even if the SSL-linear AAI is trained only on the MNGU0 speaker's voice, the model should project speech from other speakers to the same articulatory space. We provide empirical evidence to support this hypothesis in a later section (\S \ref{sec/spkid}), where we show that the inference remains consistent after converting between different voice identities.

\subsubsection{Source Features}
\label{methods:source}

Though EMA has a full descriptive capacity of the place of articulation, it lacks source information generated by the glottal excitation, which is crucial to implementing the manner of articulation and expressing the prosody of speech. Therefore, we include pitch (or fundamental frequency, f0) and loudness features to represent the source features \cite{choi2021nansy, choi2022nansy++, yang2024streamvc}. The loudness feature also informs non-larynx constriction, which is important for voiced fricatives such as ``z" and ``v". We use CREPE \cite{kim2018crepe} to infer pitch from speech, and loudness is measured by the average of absolute magnitudes of waves for every 20 ms. 
Together with the EMA from AAI, we referred to these features as ``articulatory features" that have 14 channels (12 EMA + 2 source) and a 50 Hz sampling rate.

\subsubsection{Speaker Identity} 

Since we use a template space for the vocal tract, the articulatory features lack information about the individual structures of the vocal tract anatomy. However, this structural information is an important determinant of the voice texture of an individual speaker, which is crucial in defining the speaker's identity \cite{netzorg2024percetualqaulity}. For example, the vocal tract length is known to be correlated with gender and age in voice. Note that our definition of the speaker identity does not include information about dialect or accent which is actually aimed to be disentangled from the speaker identity. 
Here, we compensate for this missing information with a separate speaker identity encoder which is jointly trained with the vocoder to extract the speaker-specific texture information.

To this end, we propose a simple yet effective speaker encoder, which minimizes the trainable portion of the model. Based on the observation by \cite{chen2022spkinfo, fan2020exploring} that the speaker information is largely concentrated in the CNN outputs of speech SSL models, the encoder consists of the frozen CNN extractor from WavLM Large followed by a weighted pooling layer and a learnable feedforward network (FFN). The pooling layer weighted-averages the acoustic features from WavLM CNN across frames, where the weight is given by the periodicity inferred from CREPE. This allows more attention to the periodic signals which may encode more information about voice texture than non-periodic portion of the input. Then, the FFN transforms the averaged features to a speaker embedding with 64 channels. This speaker identity encoding is indispensable to fully represent multi-speaker speech.

\subsection{Articulatory Synthesis}

We adopt HiFi-GAN as the vocoder for articulatory synthesis \cite{kong2020hifi, wu2022artsynth, polyak2021resynthesis}. The vocoder is trained to synthesize speech audio with a 16K Hz sampling rate from the articulatory features. To condition on the speaker embedding, we apply FiLM \cite{perez2018film} to each convolution module in the HiFi-GAN architecture, which modulates the output channels of each module. We adopt the same loss functions as \cite{kong2020hifi}: mel spectrogram loss for reconstruction and multi-period and multi-scale discriminator loss for GAN training. 

\subsection{Dataset}

For training the vocoder and speaker encoder, we use LibriTTS-R \cite{koizumi2023librittsr}, an enhanced version of LibriTTS \cite{zen2019libritts}. The dataset is comprised of 585 hours of reading audiobooks (555 hours for training). The original 24K Hz audio is downsampled to 16K Hz. We use VCTK \cite{veaux2017vctk} to further evaluate the generalizability of the model to a broader range of speakers and accents. The entire VCTK dataset is unseen during training and only used for evaluation. 

Note that the FFN in the speaker identity encoding and the HiFi-GAN vocoder are the only trainable modules (orange modules in Fig. \ref{fig:pipeline}), and the rest of the pipeline remains fixed while training. More details of implementation and training are in Appendix B.1-7.

\section{Experiments \& Results}
\label{sec/result_exp}
We evaluate the proposed SPARC in different aspects to validate each part of the framework -- universality of articulatory encoding (\ref{sec/univ_aai}), information conservation in encoding-decoding (\ref{sec/resynth}), multilingual generalization (\ref{sec/multiling}), and disentanglement of speaker identity encoding (\ref{sec/spkid}).

\subsection{Universality of Single-Speaker Articulatory Encoding}
\label{sec/univ_aai}

\begin{figure}[t]
\centering
\includegraphics[width=1\linewidth]{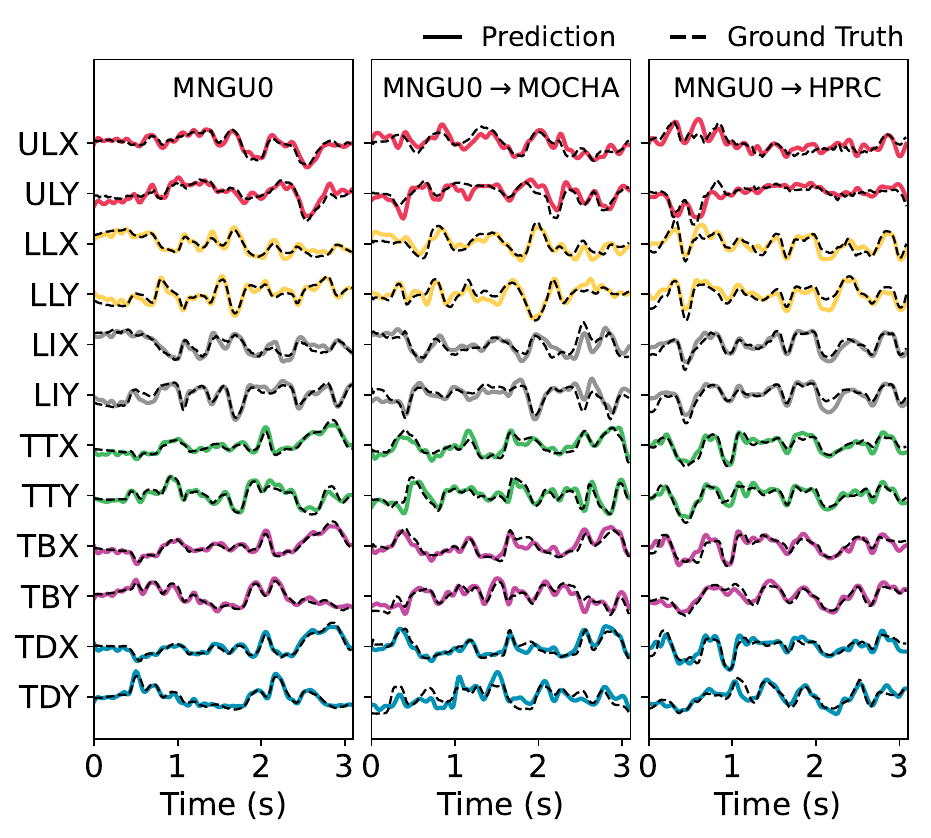}
\vspace{-15pt}
\caption{Examples of SSL-linear prediction on MNGU0 (left), transformed prediction from MNGU0 to a female MOCHA speaker (middle) and to a male HPRC speaker (right). Predictions are denoted with the colored lines and ground truths are denoted with the black dotted lines. }
\label{fig:univ_aai}
\vspace{-5pt}
\end{figure}

As we claim a single-speaker AAI can be used for universal articulatory encoding, we compare our model with the existing state-of-the-art (SOTA) multi-speaker articulatory encoding systems that utilize tract variables \cite{wu2023speakerind, siriwardena2023secret}. The tract variables are a set of articulatory parameters that can be derived from EMA and are known to be more regularized across speakers \cite{mcgowan1994recovering}. Table \ref{table:multi_aai} denotes the performance of the tract-variable-based AAI systems and ours on two multi-speaker EMA datasets, MOCHA \cite{mochatimit} and HPRC \cite{hprc}. MOCHA includes 7 speakers with 27 minutes of data per speaker on average, and HPRC includes 8 speakers with 59 minutes of data per speaker on average. The performance is evaluated by the Pearson correlation coefficient (PCC), where results besides ours are retrieved from reference papers. (95\% confidence interval is denoted for our case.) Since our AAI is trained on a single-speaker (MNGU0) articulatory space, the model outputs are not directly comparable to the EMA from MOCHA-TIMIT or HPRC. Therefore, we fit a linear model to spatially align MNGU0's articulatory space to another speaker's articulatory space to measure the correlation \cite{cho2023self}. (See Appendix A.2. for details.) As a result, our single-speaker AAI system shows correlations higher in MOCHA and slightly lower in HPRC compared to Wu et al. \cite{wu2023speakerind}, and slightly higher in HPRC compared to Siriwardena and Espy-Wilson \cite{siriwardena2023secret}. Note that this is not a fair comparison since the scores are measured on different variables. However, this system-wise comparison suggests that the single-speaker approach can yield a similar level of consistency across multiple speakers. Moreover, our approach only fits a linear model, thus it is likely to be more generalizable than the previous systems which use non-linear, deep recurrent neural networks. Fig. \ref{fig:univ_aai} demonstrates prediction examples of our AAI model. The left panel shows a near-perfect prediction even with a simple linear mapping from WavLM, and even after the affine transformation from MNGU0 space to other speakers, the predictions show high correlations with the ground truths (middle and right panel). The prediction performance on MNGU0 shows \(0.878 \pm 0.012\) average correlation, and additional analyses on AAI are reported in Appendix A.3-5. 

\begin{table}[t]
\begin{center}
\caption{Comparison of Multi-speaker AAI systems. Performance is measured as Pearson correlation coefficient (PCC).}
\scriptsize
\begin{tabular}{|l|c|cc|}
\hline
\multicolumn{1}{|c|}{\multirow{2}{*}{System}} & \multirow{2}{*}{Model}                                                         & \multicolumn{2}{c|}{Dataset}                                      \\ \cline{3-4} 
\multicolumn{1}{|c|}{}                           &                                                                                & \multicolumn{1}{c|}{MOCHA }               & HPRC                   \\ \hline
\multirow{3}{*}{Tract Variable}                  & Wu et al. \cite{wu2023speakerind}                             & \multicolumn{1}{c|}{0.678}               & 0.784                  \\ \cline{2-4} 
                                                 & \multicolumn{1}{l|}{Siriwardena \&}                                            & \multicolumn{1}{c|}{\multirow{2}{*}{--}} & \multirow{2}{*}{0.757} \\
                                                 & \multicolumn{1}{r|}{Espy-Wilson \cite{siriwardena2023secret}} & \multicolumn{1}{c|}{}                    &                        \\ \hline
\multicolumn{1}{|c|}{Single-speaker}             & Ours                                                                           & \multicolumn{1}{c|}{\(0.739 \pm 0.017\)} & \(0.761 \pm 0.012\)   \\ \hline
\end{tabular}
\label{table:multi_aai}
\vspace{-10pt}
\end{center}

\end{table}

\subsection{Performance of Resynthesis by SPARC}
\label{sec/resynth}

\begin{table}[t]

\begin{center}
\caption{Performance comparison of ground truth speech (GT) and resynthesized speech by SPARC (Resynth)}
\scriptsize
\begin{tabular}{|c|c|cc|cc|}
\hline
\multirow{2}{*}{Dataset}    & \multirow{2}{*}{Input} & \multicolumn{2}{c|}{Intelligibility}     & \multicolumn{2}{c|}{Quality}            \\ \cline{3-6} 
                            &                        & \multicolumn{1}{c|}{WER\(\downarrow\)} & CER\(\downarrow\) & \multicolumn{1}{c|}{MOS\(\uparrow\)}          & \multicolumn{1}{c|}{UTMOS\(\uparrow\)} \\ \hline
\multirow{2}{*}{LibriTTS-R} & GT                     & \multicolumn{1}{c|}{4.22}              & 2.24              & \multicolumn{1}{c|}{\(4.01 \pm 0.16\)}        & \multicolumn{1}{c|}{\(4.10 \pm 0.58\)} \\ \cline{2-6} 
                            & Resynth                & \multicolumn{1}{c|}{5.43}              & 2.90              & \multicolumn{1}{c|}{\(3.82 \pm 0.20\)}        & \multicolumn{1}{c|}{\(4.12 \pm 0.58\)} \\ \hline
\multirow{2}{*}{VCTK}       & GT                     & \multicolumn{1}{c|}{1.66}              & 0.74              & \multicolumn{1}{c|}{\(3.65 \pm 0.16\)}        & \multicolumn{1}{c|}{\(3.84 \pm 0.54\)} \\ \cline{2-6} 
                            & Resynth                & \multicolumn{1}{c|}{3.73}              & 1.96              & \multicolumn{1}{c|}{\(3.67 \pm 0.17\)}        & \multicolumn{1}{c|}{\(3.95 \pm 0.52\)} \\ \hline
\end{tabular}
\label{table:performance}
\end{center}
\end{table}

We measure the intelligibility and quality of resynthesized speech audio to evaluate how well information is preserved in the encoding-decoding process.  We use an out-of-the-box automatic speech recognition (ASR) model, Whisper (``openai/whisper-large-v3") \cite{radford2023robust}, to evaluate the word error rate (WER) and character error rate (CER) of resynthesized speech.\footnote{The transcription is normalized before measuring the error rates.} The quality of the audio is measured by a human-evaluated subjective metric, mean opinion score (MOS), and a machine-evaluated MOS, UTMOS \cite{saeki2022utmos}. We evaluate both ground truth speech and resynthesized speech on the test-clean subset of LibriTTS-R which includes 8.56 hours from 39 unseen speakers. To further evaluate the generalizability of the model, we evaluate the model on VCTK dataset that includes 107 speakers with 500 utterances in total. Table \ref{table:performance} summarizes the performance in comparison with the ground truth speech. For the quality metrics, we reported 95\% confidence intervals. When tested on LibriTTS-R, the ASR on the resynthesized speech shows high intelligibility with WER of 5.43\% and CER of 2.90\%, which is marginally different from those of the ground truth. Moreover, our articulatory synthesis generates natural speech sounds showing decent quality with MOS of 3.82 which is a 0.19 decrease from the ground truth, and with UTOMOS of 4.12 which is on par with the ground truth.

Furthermore, the results on VCTK also demonstrate high intelligibility with WER of 3.73\% and CER of 1.96\%. Though the gap from the ground truth is larger than LibriTTS-R, this is a remarkable level of performance given that the model is not exposed to the entire dataset during training. As LibriTTS-R has cleaner audio than VCTK, the model demonstrates some speech enhancement capacity which results in a higher MOS and UTMOS than the ground truth. While the speakers in LibriTTS-R are concentrated on American English, VCTK includes a variety of accents, primarily focusing on English speakers with different regional accents from the United Kingdom. Therefore, the high performance demonstrated on VCTK suggests that SPARC can robustly generalize to unseen speakers and accents.

\subsection{Multilingual Generalization}
\label{sec/multiling}
\begin{table*}[t]

\begin{center}
\caption{Performance comparison on multilingual speech data. Each metric is evaluated on ground truth speech (GT) and resynthesized speech by English-only trained model (EN-R) or multilingual model (ML-R).}
\scriptsize
\begin{tabular}{|c|ccc|ccc|ccc|}
\hline
\multirow{2}{*}{Language} & \multicolumn{3}{c|}{WER\(\downarrow\)}           & \multicolumn{3}{c|}{CER\(\downarrow\)}              & \multicolumn{3}{c|}{UTMOS\(\uparrow\)}                           \\ \cline{2-10} 
    & \multicolumn{1}{c|}{GT} & \multicolumn{1}{c|}{EN-R} & ML-R & \multicolumn{1}{c|}{GT} & \multicolumn{1}{c|}{EN-R} & ML-R  &  \multicolumn{1}{c|}{GT} & \multicolumn{1}{c|}{EN-R} & ML-R \\ \hline
German & \multicolumn{1}{c|}{5.23}   & \multicolumn{1}{c|}{15.32}   & \multicolumn{1}{c|}{9.70}   & \multicolumn{1}{c|}{1.89}   & \multicolumn{1}{c|}{6.66}   & \multicolumn{1}{c|}{3.95}  & \multicolumn{1}{c|}{\(2.96 \pm 0.83\)}   & \multicolumn{1}{c|}{\(2.87 \pm 0.58\)}   & \multicolumn{1}{c|}{\(2.83 \pm 0.72\)} \\ 
Dutch & \multicolumn{1}{c|}{9.90}   & \multicolumn{1}{c|}{20.20}   & \multicolumn{1}{c|}{15.39}   & \multicolumn{1}{c|}{2.82}   & \multicolumn{1}{c|}{8.83}   & \multicolumn{1}{c|}{5.70}  & \multicolumn{1}{c|}{\(2.91 \pm 0.82\)}   & \multicolumn{1}{c|}{\(2.89 \pm 0.65\)}   & \multicolumn{1}{c|}{\(2.77 \pm 0.79\)} \\ 
Portuguese & \multicolumn{1}{c|}{7.10}   & \multicolumn{1}{c|}{17.81}   & \multicolumn{1}{c|}{12.80}   & \multicolumn{1}{c|}{2.19}   & \multicolumn{1}{c|}{8.50}   & \multicolumn{1}{c|}{5.21} & \multicolumn{1}{c|}{\(2.63 \pm 1.34\)}   & \multicolumn{1}{c|}{\(2.81 \pm 0.67\)}   & \multicolumn{1}{c|}{ \(2.56 \pm 0.92\)} \\ 
Italian & \multicolumn{1}{c|}{9.63}   & \multicolumn{1}{c|}{22.09}   & \multicolumn{1}{c|}{15.53}   & \multicolumn{1}{c|}{2.43}   & \multicolumn{1}{c|}{7.40}   & \multicolumn{1}{c|}{4.98}  & \multicolumn{1}{c|}{ \(2.78 \pm 1.26\)}   & \multicolumn{1}{c|}{ \(2.82 \pm 0.61\)}   & \multicolumn{1}{c|}{\(2.67 \pm 0.82\)}\\ 
Polish & \multicolumn{1}{c|}{4.19}   & \multicolumn{1}{c|}{14.79}   & \multicolumn{1}{c|}{9.89}   & \multicolumn{1}{c|}{0.94}   & \multicolumn{1}{c|}{4.63}   & \multicolumn{1}{c|}{2.87}  & \multicolumn{1}{c|}{\(3.09 \pm 0.76\)}   & \multicolumn{1}{c|}{\(3.00 \pm 0.63\)}   & \multicolumn{1}{c|}{\(2.95 \pm 0.76\)}\\ 
Spanish & \multicolumn{1}{c|}{3.96}   & \multicolumn{1}{c|}{10.74}   & \multicolumn{1}{c|}{6.79}   & \multicolumn{1}{c|}{1.26}   & \multicolumn{1}{c|}{4.73}   & \multicolumn{1}{c|}{2.38}  & \multicolumn{1}{c|}{\(2.93 \pm 0.86\)}   & \multicolumn{1}{c|}{ \(2.88 \pm 0.66\)}   & \multicolumn{1}{c|}{\(2.72 \pm 0.81\)}\\ 
French & \multicolumn{1}{c|}{4.62}   & \multicolumn{1}{c|}{16.67}   & \multicolumn{1}{c|}{8.04}   & \multicolumn{1}{c|}{2.65}   & \multicolumn{1}{c|}{9.90}   & \multicolumn{1}{c|}{4.54}  & \multicolumn{1}{c|}{\(2.79 \pm 0.93\)}   & \multicolumn{1}{c|}{\(2.80 \pm 0.65\)}   & \multicolumn{1}{c|}{\(2.69 \pm 0.79\)}\\ 
Korean & \multicolumn{1}{c|}{6.31}   & \multicolumn{1}{c|}{15.88}   & \multicolumn{1}{c|}{13.66}   & \multicolumn{1}{c|}{1.01}   & \multicolumn{1}{c|}{6.72}   & \multicolumn{1}{c|}{4.96}  & \multicolumn{1}{c|}{\(3.69 \pm 0.63\)}   & \multicolumn{1}{c|}{\(3.47 \pm 0.80\)}   & \multicolumn{1}{c|}{\(2.91 \pm 0.95\)}\\ 
Japanese & \multicolumn{1}{c|}{--}   & \multicolumn{1}{c|}{--}   & \multicolumn{1}{c|}{--}   & \multicolumn{1}{c|}{5.65}   & \multicolumn{1}{c|}{8.45}   & \multicolumn{1}{c|}{8.16}  & \multicolumn{1}{c|}{\(3.53 \pm 0.80\)}   & \multicolumn{1}{c|}{\(3.49 \pm 0.77\)}   & \multicolumn{1}{c|}{\(3.08 \pm 0.83\)}\\ 
Chinese & \multicolumn{1}{c|}{--}   & \multicolumn{1}{c|}{--}   & \multicolumn{1}{c|}{--}   & \multicolumn{1}{c|}{10.69}   & \multicolumn{1}{c|}{31.27}   & \multicolumn{1}{c|}{23.28}  & \multicolumn{1}{c|}{\(2.66 \pm 1.21\)}   & \multicolumn{1}{c|}{ \(3.06 \pm 0.92\)}   & \multicolumn{1}{c|}{\(2.71 \pm 1.12\)} \\ \hline
Average & \multicolumn{1}{c|}{6.37}   & \multicolumn{1}{c|}{16.69}   & \multicolumn{1}{c|}{11.48}   & \multicolumn{1}{c|}{3.15}   & \multicolumn{1}{c|}{9.71}   & \multicolumn{1}{c|}{6.60}  & \multicolumn{1}{c|}{\(3.00 \pm 0.66\)}   & \multicolumn{1}{c|}{\(3.01 \pm 0.49\)}   & \multicolumn{1}{c|}{\(2.79 \pm 0.29\)} \\ 
w/o Chinese & \multicolumn{1}{c|}{6.37}   & \multicolumn{1}{c|}{16.69}   & \multicolumn{1}{c|}{11.48}   & \multicolumn{1}{c|}{2.32}   & \multicolumn{1}{c|}{7.31}   & \multicolumn{1}{c|}{4.75}   & \multicolumn{1}{c|}{--}   & \multicolumn{1}{c|}{--}   & \multicolumn{1}{c|}{--} \\ \hline
\end{tabular}
\label{table:multilingua}
\end{center}
\end{table*}

We evaluate SPARC on speech corpora from other languages: 7 European languages (German, Dutch, Portuguese, Italian, Polish, Spanish, French) from multilingual LibriSpeech \cite{pratap2020mls} and 3 East Asian languages (Korean, Japanese, and Chinese (Mandarin)) from KSS \cite{park2018kss}, JVS \cite{takamichi2019jvs}, and AISHELL \cite{shi2020aishell}, respectively. Except for KSS, all corpora include multiple speakers. We randomly sampled 200 utterances from the test split of each corpus to evaluate the resynthesized speech.\footnote{We excluded samples with numerals to avoid converting Arabic numerals. } Table \ref{table:multilingua} compares the intelligibility (WER and CER using Whisper) and quality (UTMOS). As no interword spacing is used for Japanese and Chinese orthographic systems, words in these languages are not separated as straightforwardly as other languages. Therefore, we omit the WER for these languages. We also evaluate the performance of a multilingual version of the model that is obtained by fine-tuning to the multilingual data (denoted as ML-R in Table \ref{table:multilingua}). The original English-only trained model is denoted as EN-R. See Appendix B.8. for details of the fine-tuning procedure.

As shown in Table \ref{table:multilingua}, the resynthesis by English-only model (EN-R) achieves average WER of 16.69 excluding Japanese and Chinese, average CER of 9.71. Roughly, if the conflated Chinese CER is excluded, the EN-R preserves a fair amount of information given that 93\% of characters are correctly recognized. Despite the huge gap from the ground truth, this is remarkable generalizability given both encoding and decoding procedures have only seen English in training. When fine-tuned (ML-R), the average WER and CER are cut down to 11.48\% and 6.60\%, respectively. This indicates that some portion of the errors is induced by the out-of-domain application of the vocoder, thus that fine-tuning is able to yield a huge gain in performance. Yet, there are unresolved gaps from the ground truth, which would be attributed to the fact that we are only tuning the vocoder and keeping the encoder part intact. This is also aligned with the observation that there is a slight language bias in the articulatory representation in SSL \cite{cho2023self}. In terms of synthesis quality, we find a minimal difference between ground truth (\(3.00 \pm 0.66\)) and resynthesized audio from English-only trained model (\(3.01 \pm 0.49\)), and the quality slightly drops to \(2.79 \pm 0.29\) when the model is fine-tuned to multilingual dataset. (The scores are averaged across languages and the ranges denote 95\% confidence intervals.) The drop in quality is likely due to the general inferiority in the quality of multilingual datasets compared to LibriTTS-R.

\subsection{Speaker Recognition and Voice Control by Speaker Identity}
\label{sec/spkid}

\begin{table}[t]
\begin{center}
\caption{Few-Shot Speaker identification accuracy (SID ACC). For VCTK, the accuracy of full-train data template embeddings is denoted in parentheses.}
\begin{tabular}{|c|cc|}
\hline
\multicolumn{1}{|c|}{\multirow{2}{*}{Model}} & \multicolumn{2}{c|}{SID ACC (\%)\(\uparrow\)}                           \\ \cline{2-3} 
\multicolumn{1}{|c|}{}                       & \multicolumn{1}{c|}{LibriTTS-R} & \multicolumn{1}{c|}{VCTK} \\ \hline
x-vector                                     & \multicolumn{1}{c|}{91.7}       & \multicolumn{1}{c|}{83.4 (88.2)}               \\ \hline
r-vector                                    & \multicolumn{1}{c|}{98.3}       & \multicolumn{1}{c|}{99.8 (99.8)}               \\ \hline
Ours                       & \multicolumn{1}{c|}{95.5}       & \multicolumn{1}{c|}{92.2 (94.6)}              \\ \hline
\end{tabular}
\vspace{-10pt}
\label{table:timbre_sid}
\end{center}
\end{table}

The speaker identity encoding informs voice texture of speech, which is especially important as our articulatory features are speaker agnostic. On one hand, this suggests that the model can learn a speaker embedding that is disentangled from individual tendencies in articulation, or accents. Here, we evaluate this claim by experiments on few-shot speaker identification and zero-shot voice conversion. 

First, we build a few-shot, learning-free speaker identification (SID) by comparing the similarity between the speaker embeddings. For each test speaker, 10 first clips in the dataset are concatenated and then the speaker embedding is extracted to serve as a template embedding for the speaker. Then among potential test speakers, the identity is predicted by choosing the speaker with maximal similarity between the template and tested speaker embedding. For LibriTTS-R, we use speakers in the test-clean set with at least 10 clips, leaving 36 speakers. For VCTK, we use the train set clips to create the template embeddings for 107 test speakers. Additionally, we also evaluate SID accuracy using all data in the train set for creating the templates, which is denoted in parentheses in Table \ref{table:timbre_sid}.

\begin{figure}[t]
\centering
\includegraphics[width=1\linewidth]{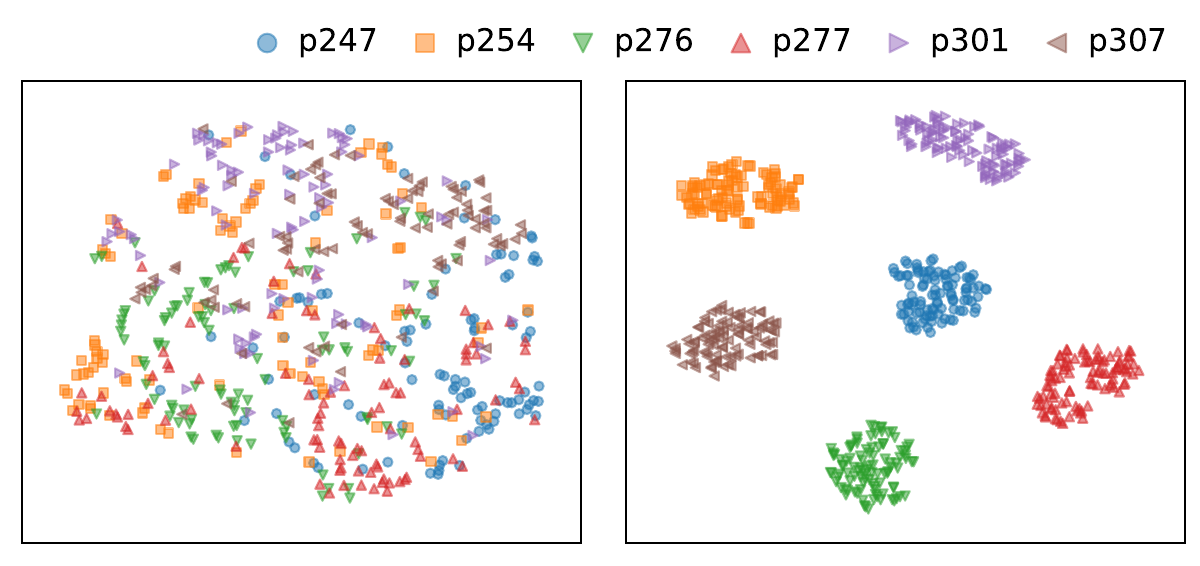}
\vspace{-15pt}
\caption{Visualization by T-SNE of utterance-wise averaged articulations (left) and speaker embeddings (right) from 6 different VCTK speakers. The perplexity is set as 10.}
\label{fig:spk_tsne}
\end{figure}

Our speaker encoding shows high discriminability in the few-shot SID, achieving 94.7\% accuracy in LibriTTS-R and 92.2 \% accuracy in VCTK test sets. When all utterances in the train set are used, the accuracy for VCTK reaches 94.6 \%. Our embedding outperforms the widely adopted speaker embedding, x-vector \cite{snyder2018x} by a large margin, especially in VCTK which is a harder case with a larger number of speakers (Table \ref{table:timbre_sid}).\footnote{We retrieved checkpoints from https://huggingface.co/speechbrain/spkrec-xvect-voxceleb.} As a sanity check for the proposed SID task, we evaluate the accuracy with the current state-of-the-art speaker embedding, ResNet based r-vector by WeSpeaker \cite{wang2023wespeaker}, which shows a near-perfect SID accuracy.\footnote{We use "resnet221\_LM" checkpoint from https://github.com/wenet-e2e/wespeaker.} However, the r-vector embedding is extracted from very deep ResNet with 221 layers, which is way larger than ours and x-vector, therefore the model may incorporate more speaker information than voice texture (e.g., dialect and accent). Both x-vector and r-vector models are trained on VoxCeleb corpus \cite{nagrani2017voxceleb, chung2018voxceleb2, nagrani2020voxceleb}. Overall, this suggests that our speaker encoding can extract highly discriminable features for speaker identification.

We further evaluate the validity of our identity encoding by zero-shot voice cloning application. We selected 6 speakers from VCTK, where their mean pitch values are equidistant in the range from 80 Hz to 230 Hz. Fig. \ref{fig:spk_tsne} visualizes the speaker embedding of the selected speakers, demonstrating that each speaker's utterances are clustered distinctively. On the contrary, no speaker-relevant cluster is identifiable in the manifold of utterance-wise averaged articulations. For each target speaker, the first 10 clips in the training set are concatenated and then the speaker embedding is extracted. Then, the voice from the source speaker is converted by switching speaker embedding in the synthesis. Additionally, the pitch range of the source speaker is adjusted to match that of the target speaker, by z-scoring the source pitch trace; and shifting and rescaling to match the center and scale of the target pitch range (``P-rescale"). We convert all audio clips in the test-clean set of LibriTTS-R with more than 2 seconds of duration, to each of 6 VCTK target speakers.

\begin{table*}[t]
\begin{center}
\caption{Performance comparison of Zero-Shot Voice conversion. The SID accuracy using our speaker encoding is denoted in parenthesis. }
\label{table:vc}
\begin{tabular}{|c|c|cccc|c|c|c|}
\hline
\multirow{2}{*}{Model} & \multirow{2}{*}{Target Spk} & \multicolumn{4}{c|}{Coding-Recoding Similarity (PCC)}                                                                                                      & \multirow{2}{*}{SID\(\uparrow\)} & \multirow{2}{*}{WER\(\downarrow\)} & \multirow{2}{*}{UTMOS\(\uparrow\)} \\ \cline{3-6}
                       &                             & \multicolumn{1}{c|}{Articulation\(\uparrow\)} & \multicolumn{1}{c|}{Pitch\(\uparrow\)}   & \multicolumn{1}{c|}{Loudness\(\uparrow\)} & Speaker\(\uparrow\)  &                                  &                                    &                                    \\ \hline
FreeVC                & VCTK                        & \multicolumn{1}{c|}{\(0.923 \pm 0.021\)}      & \multicolumn{1}{c|}{\(0.881 \pm 0.165\)} & \multicolumn{1}{c|}{\(0.881 \pm 0.062\)}  & \(0.839 \pm 0.109\) & 92.5 (81.4)                             & 4.00                               & \(4.05 \pm 0.51\)                  \\ \hline
QuickVC                & VCTK                        & \multicolumn{1}{c|}{\(0.930 \pm 0.021\)}      & \multicolumn{1}{c|}{\(0.868 \pm 0.172\)} & \multicolumn{1}{c|}{\(0.903 \pm 0.063\)}  & \(0.971 \pm 0.105\) & 65.0 (66.7)                             & 4.17                                & \(4.22 \pm 0.36\)                  \\ \hline
Ours                   & VCTK                        & \multicolumn{1}{c|}{\(0.944 \pm 0.015\)}      & \multicolumn{1}{c|}{\(0.944 \pm 0.093\)} & \multicolumn{1}{c|}{\(0.938 \pm 0.037\)}  & \(0.879 \pm 0.069\) & 94.6                             & 4.83                               & \(3.83 \pm 0.58\)                  \\
w/o P-rescale          & VCTK                        & \multicolumn{1}{c|}{\(0.943 \pm 0.016\)}      & \multicolumn{1}{c|}{\(0.894 \pm 0.150\)} & \multicolumn{1}{c|}{\(0.929 \pm 0.045\)}  & \(0.791 \pm 0.100\) & 73.9                             & 4.82                               & \(3.77 \pm 0.78\)                  \\ \hline
Ours-Resynth           & Self                        & \multicolumn{1}{c|}{\(0.954 \pm 0.014\)}      & \multicolumn{1}{c|}{\(0.968 \pm 0.109\)} & \multicolumn{1}{c|}{\(0.936 \pm 0.077\)}  & \(0.962 \pm 0.038\) &      --                          & 4.40                               & \(4.00 \pm 0.59\)                  \\ \hline
\end{tabular}
\end{center}
\end{table*}

We evaluate ``coding-recoding similarity" of the voice-converted speech. The coding-recoding similarity measures the correlation between the speech coding of original speech and that of synthesized speech. Note that our coding pipeline is also an analytic measurement that provides articulatory and source features of speech. Therefore, the metric indicates how much the articulation and source information is preserved in the voice conversion and speech synthesis process. For articulation, we measure the Pearson correlation of each of 12 channels and then average them, and for pitch and loudness, the correlation is separately reported. For speaker identity, the cosine similarity is measured between the conditioning target embedding and the embedding extracted from synthesized speech. Unlike articulatory and source features, speaker embedding is arbitrary and not interpretable. Therefore, we evaluate the accuracy in SID task, the same discriminability task designed above for VCTK. Since the task requires to pinpoint the target speaker from 107 speakers, the SID accuracy can inform how closely the voice-converted speech matches the target speaker while being distinct from others. Lastly, we measure WER and UTMOS to evaluate the intelligibility and quality of the converted audio. As baselines, we evaluate existing voice conversion models: FreeVC \cite{li2023freevc} and QuickVC \cite{guo2023quickvc}. Both models encode speech contents using speech SSL models and have separate speaker encoders jointly trained with synthesizers. The former is more comparable to ours as the content encoder is also based on  WavLM. However, none of the baseline models impose any grounded nor interpretable structures on the speech codes. We apply our framework to analyze articulatory and source features of the speech synthesized by baselines, while the correlations of speaker embeddings and SID scores are evaluated using their own speaker encoders. We also denote the SID accuracies using our speaker encoder in parentheses (Table. \ref{table:vc}). Furthermore, we evaluate the coding-recoding similarity of the self-targeted resynthesized speech (``Ours-Resynth" in Table. \ref{table:vc}), to demonstrate consistency of the speech content in synthesis without voice conversion, which represents the upper bound of the scores. To avoid confusion, we do not report SID for resynthesis since the speaker identity is not imposed to be VCTK speakers. The results are summarized in Table. \ref{table:vc} with 95\% confidence intervals if applicable.

As a result, the voice-converted speech by our model can be accurately identified as the target speaker among 107 VCTK speakers, achieving 94.6 \% accuracy (Table. \ref{table:vc}). Though the similarity in speaker embedding is lower in voice conversion than in the self-targeted resynthesis (0.879 and 0.962, respectively), the SID result suggests that this level of similarity is high enough to be discriminable from 107 unseen speakers. Furthermore, our approach shows higher SID accuracy than both of the baseline models. FreeVC shows inferior scores in both speaker similarity and SID scores. QuickVC shows higher speaker similarity than ours but shows significantly lower SID accruacy as 65\% (66.7\% using our speaker encoding). The latter case indicates that their speaker embedding is relatively simple to be more consistent through voice conversion, but lacks specificity to discriminate different speakers.

The SID accuracy and the coding-recoding similarity of pitch and speaker identity significantly drop when we ablate the pitch rescaling (``w/o P-rescale" in Table. \ref{table:vc}), indicating an interplay between pitch range and speaker identity encoding. This may be induced by the natural correlation between speaker identity and pitch range, which is likely to be confounded by biological sex. However, the coding-recoding similarity of articulation and loudness remains intact, showing a marginal difference from the pitch-rescaled voice conversion. 

The high coding-recoding similarity in articulation suggests that the articulation is well disentangled from speaker identity and pitch, which indicates that accent is preserved after the voice conversion. The correlation is as high as 0.944 and is marginally different from the self-targeted resynthesis, 0.954. This also indicates the stability of the coding by SPARC. This result further supports the claim that the proposed AAI is agnostic to speaker identity, allowing highly consistent inference even with different voice identities. Compared to the baseline models, SPARC demonstrates superior similarities, especially in pitch, suggesting our model better disentangles pitch from the other features. Regarding the intelligibility, WER slightly increases after the voice conversion, but the difference is marginal and the speech remains highly intelligible with WER of 4.83 \%. Both baselines show lower WERs than our model (FreeVC: 4.00 \%, and QuickVC: 4.17 \%) indicating a potential internal process that might improve intelligibility by modifying the articulatory features of the original speech. Finally, the voice conversion shows some degraded UTMOS as 3.83, 0.17 decrease compared to the resynthesis, and QuickVC shows the highest UTMOS as 4.22. This suggests that our model has a potential trade-off between disentanglement and audio quality of the synthesized audio. However, this degrades in naturalness may be confounded by a natural correlation between accents and voice textures, so that some voice conversion cases may result in rare combinations of accent and speaker identity, harming the UTMOS score. To sum up, these results indicate that SPARC is a disentangled and stable coding system of speech.

\section{Demonstration of Interpretability and Controllability of SPARC}

The proposed articulatory coding is highly interpretable and controllable, which is very unique compared to previous neural coding approaches. Interpretability is naturally granted as each feature represents a physical articulator on the vocal tract. Furthermore, the articulatory features can be controlled with the same principles that govern speech production. The control or modification of articulatory features can be played back by the vocoder, providing an audible feedback of the articulatory control. Together, SPARC can function as a physical simulation of vocal tract articulation \cite{maeda1990compensatory, story2005parametric, birkholz2013prev_artsynth, birkholz2017manipulation, gao2024copysynthesis}. Here, we demonstrate these features with two example cases. 

\begin{figure}[t]
\centering
\includegraphics[width=1\linewidth]{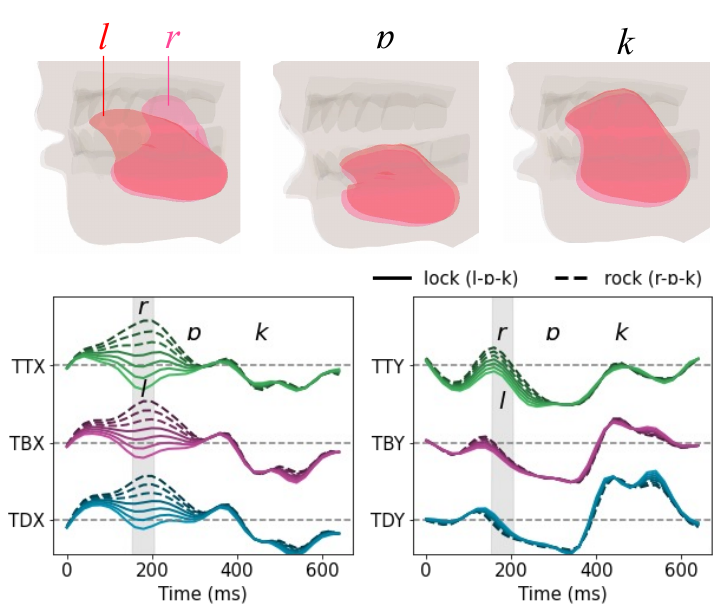}
\vspace{-10pt}
\caption{Articulatory traces encoded for speaking ``lock" and ``rock", denoted by the different line styles, "--" and "-{-}", respectively. The bottom panel shows the midsagittal displacements of TT, TB, and TD, and the top panels show snapshots of the corresponding vocal tract anatomy. In the snapshots, the vocal tract of ``lock" and ``rock" are overlaid with separate colors, orange and pink, respectively. The shaded region indicates the window of \textit{``l"} or \textit{``r"}. The color is darkened while interpolating from ``lock" to ``rock", where the line style indicates the recognized words.  }
\label{fig:interp_tongue}
\end{figure}

\subsection{Case 1: Place of Articulation -- ``lock" vs ``rock"}
First of all, we demonstrate how SPARC can be used to interpret and control the place of articulation in speech. Here, we select two speech clips of speaking ``lock (\textit{l-\textipa{\textturnscripta}-k})" and ``rock (\textit{r-\textipa{\textturnscripta}-k})" as an example pair to illustrate the difference between alveolar and post-alveolar approximants. The articulatory traces of these two words extracted by SPARC are depicted in  Fig. \ref{fig:interp_tongue}, along with snapshots of the vocal tract animation. These offer an intuitive interpretation of how the vocal tract is dynamically shaped while speaking. Furthermore, this successfully highlights the known phonological distinction between these two approximants: \textit{``l"} approximates tongue to a more anterior part of the palate (alveolar) than \textit{``r"} (post-alveolar).

Furthermore, we demonstrate a control simulation by interpolating tongue articulations between the words \cite{wu2022artsynth}. The mixing factor, \(\alpha\), is applied to weigh the sum of each of the tongue articulatory traces (TT, TB, and TD), i.e., \(Art("lock") \cdot \alpha + Art("rock") \cdot (1-\alpha)\), where \(Art(\cdot)\) means articulatory encoding. The Fig. \ref{fig:interp_tongue} bottom panel shows the interpolated traces with \(\alpha \in \{1.0, 0.8, 0.6, 0.4, 0.2, 0.0, -0.2\}\), gradually changing pronunciation from ``lock" to ``rock". Then, ASR is applied to synthesized speech to determine which word is perceived, drawing a perceptual boundary at \(\alpha=0.2\), which is denoted using different line styles in Fig. \ref{fig:interp_tongue}. This demonstration shows that SPARC provides interpretable control knobs, which can be manipulated to simulate speech sounds and observe the causal interaction between articulatory control and the generated speech.  

\subsection{Case 2: Manner of Articulation -- Voice Onset Time}

\begin{figure}[t]
\centering
\includegraphics[width=0.85\linewidth]{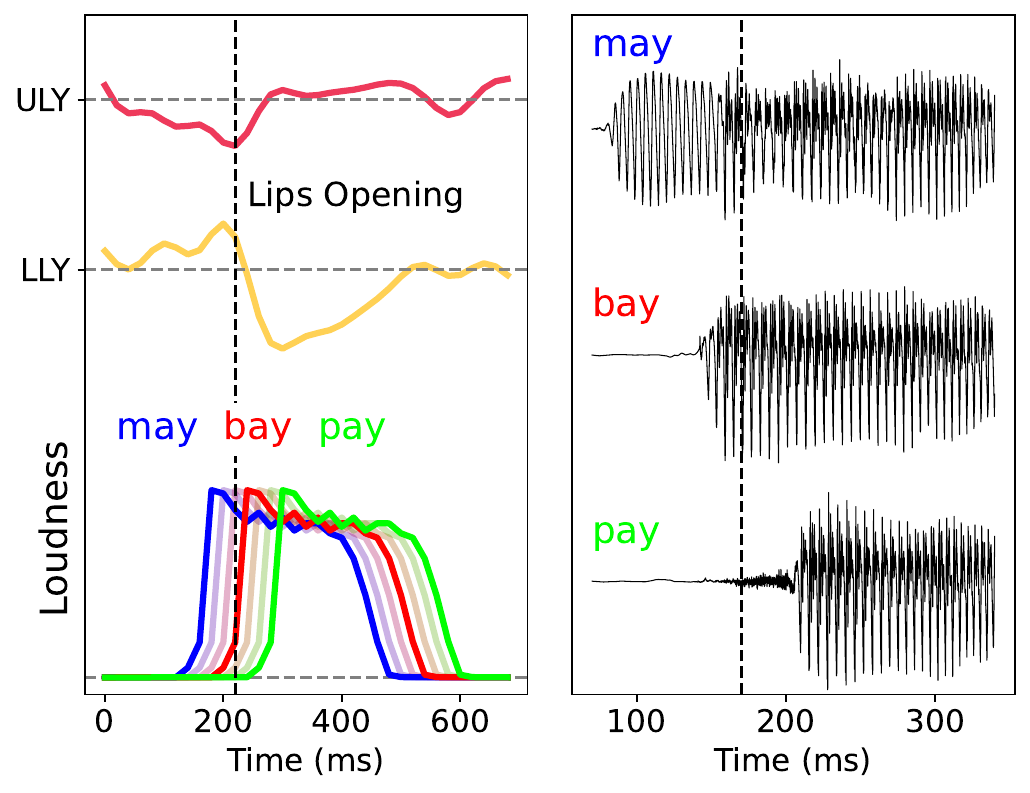}
\vspace{0pt}
\caption{Articulatory modulation example for "bay". The left panel shows the Y axis of UL and LL with the vertical dashed line indicating the beginning of the lips opening. The loudness trajectories are depicted, which are moved back and forth, while the salient red, green, and blue colors indicate the perceived plosives. The right panel shows zoom-in wave form of synthesized audio around the lips opening, showing the different voice onset times. }
\label{fig:interp_voice}
\end{figure}

The next example is about the manner of articulation, especially focused on the phonological phenomena called voice onset time (VOT). While the spatial arrangement of articulators forms filters for consonants, the manner of sound further modifies sounds into perceptually distinguishable sounds. In this example, we demonstrate that the VOT can be directly controlled by changing the timing of the rise of loudness.

Here, a speech clip of the word ``bay" is encoded by the articulatory encoder (Fig. \ref{fig:interp_voice} left). We only show the traces of lips on the Y axis as these are the key articulatory factors of labial consonants. We can clearly see the lips opening around 200 ms. Then, we shift the loudness trace back and forth and generate speech from the manipulated features. By shifting the loudness trace 60 ms earlier than the moment of lips opening, we can nasalize the sound from ``bay" to ``may" (from plosive to nasal). On the other hand, by shifting it 60 ms backward, the sound is converted to ``pay" (from voiced to voiceless plosive). We can also observe the induced VOT difference along this manipulation (Fig. \ref{fig:interp_voice} right), which is aligned with the known VOT patterns by the nasality and voicedness of sounds.

These two proof-of-concept examples demonstrate promising utilities of SPARC as a simulator of articulatory control, and as a speech analysis tool that provides a natural and intuitive phonological interpretation and intervention. 

\subsection{Potential Applications}

Based on the results, we envision several potential applications of SPARC. First, SPARC can be used as an analysis platform for investigating a phonological basis of speech without collecting high-cost articulatory data \cite{gao2024copysynthesis}. Second, the high-performance synthesizer can be used as a speech simulator, which can facilitate a control theoretic approach to speech processing or reinforcement learning of speaking agents \cite{maeda1990compensatory, patri2015optimal, birkholz2017manipulation, rakotomalala2022trajectories}. Third, the universal articulatory analysis can be utilized for a language learning tool or therapy, where visualizing the vocal tract can enhance the learning experience \cite{badin2010langlearn, suemitsu2015real, bernhardt2003speechtherapy}. Lastly, a TTS system can be built upon SPARC that synthesizes articulatory control from text or higher-order structure of speech \cite{lian2022gesture1, lian2023gesture2, cho2024sdhubert}, revealing a descriptive and interpretable relationship between text and articulation. In conclusion, SPARC holds promise for various applications in both speech science and engineering.
\section{Conclusion}

We propose a novel encoding-decoding framework of speech, \textit{SPARC}, which mimics the biophysical apparatus of speech production. Through large-scale training, our proposed SPARC framework achieves a high-performance articulatory inversion and synthesis that can generalize to unseen speakers, with a minimal loss of information compared to ground truth speech. To our knowledge, this is the first demonstration of universal articulatory synthesis that can scale up to an indefinite number of speakers. This universality is supported by a novel speaker identity encoding that embeds highly discriminable speaker-specific voice textures. Additionally, the encoded speaker embedding is effectively disentangled from articulatory features and allows accent-preserving voice conversion. Using SPARC, speech can be represented in a low-dimensional space, with each channel corresponding to an articulator in the vocal tract. This physical embodiment of speech allows for natural and intuitive control and interpretation.

In future work, we will scale the SPARC framework to incorporate expressive speech and singing. Also, we will improve the robustness of the system under noisy environments. These efforts will maximize the promising utilities of the articulatory coding of speech.

\section*{Acknowledgements}

This research is supported by the following grants to PI Anumanchipalli --- NSF award 2106928, BAIR Commons-Meta AI Research, the Hellman Fellows Program, the Rose Hills Innovator Program, UC Noyce Initiatives at UC Berkeley, and Google Research Scholar Award. We acknowledge Dr. Abdelrahman Mohamed (Rembrand) and Prof. Alan W Black (Carnegie Mellon University) for insightful discussions at various stages of this work.

\section*{Appendix A -- Additional Analysis}

\subsection*{A.1. Layer-wise AAI Performance}

\begin{figure}[h]
\centering
\includegraphics[width=0.8\linewidth]{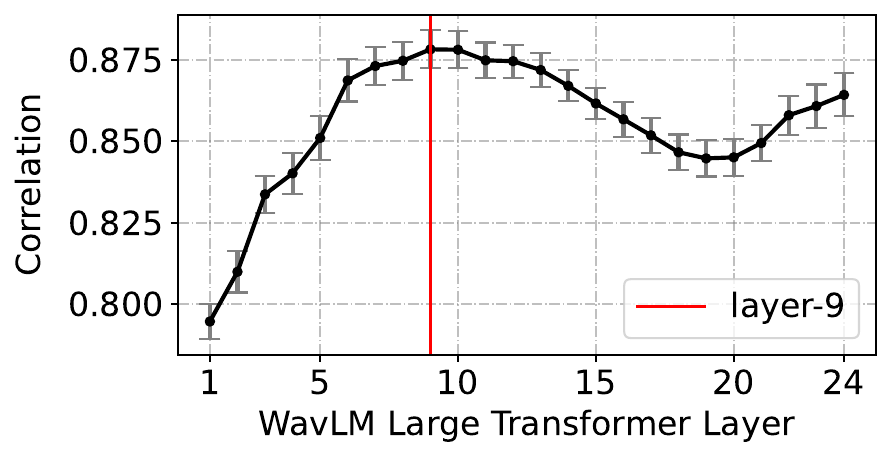}
\vspace{0pt}
\caption{Probing performance of each layer in WavLM Large on MNGU0}
\label{fig:wavlm_mngu}
\end{figure}

We selected the 9-th layer after comparing the inversion performance of each layer measured by Pearson correlation coefficients (PCC) averaged across EMA channels on 5-fold cross-validation (Fig. \ref{fig:wavlm_mngu}). For each fold, 100 test utterances are randomly held out and 95 \% confidence intervals are denoted. At the best layer (the 9th), the average correlation is \(0.878 \pm 0.012\). After the layer selection, all data from MNGU0 are used to fit the linear model. The parameters are estimated by ordinary least square (OLS) method. We find a minimal difference between with and without a regularization term (e.g., L2 penalty). Thus, we use OLS for a better simplicity.

\subsection*{A.2. Measuring Correlation Between Articulations from Different Articulatory Space.}

Since the EMA sensors are not spatially aligned across speakers, each EMA channel from different speakers is not directly comparable. As our model is using a particular participant’s (MNGU0’s) vocal tract as a template space, we find a mapping that spatially aligns MNGU0 sensors to different participants’ to calculate the correlation. Following \cite{cho2023self}, we pass speech audios from a target speaker from MOCHA-TIMIT or HPRC to our inversion model, and then a linear model is trained to predict the reference EMA of the target speaker. We used a linear regression with L1 penalty (Lasso) to impose sparsity in the weights, using \(alpha=0.01\). Scikit-learn library is used for training the model (``sklearn.linear\_model.Lasso”) \cite{scikit-learn}. This L1 penalty reveals channel-by-channel correspondence in the affine transformation (Fig. \ref{fig:coef_map}; more discussed in the later section). We split the data to 5 folds where 4 folds are used for training the linear model and 1 fold is used for evaluating the model. This is done individually for each speaker in MOCHA-TIMIT (7 speakers with 27 m of speech in average) and HPRC (7 speakers with 59 m of speech in average).

\subsection*{A.3. AAI Performance on Individual Articulator}

\begin{figure}[h]
\centering
\includegraphics[width=0.8\linewidth]{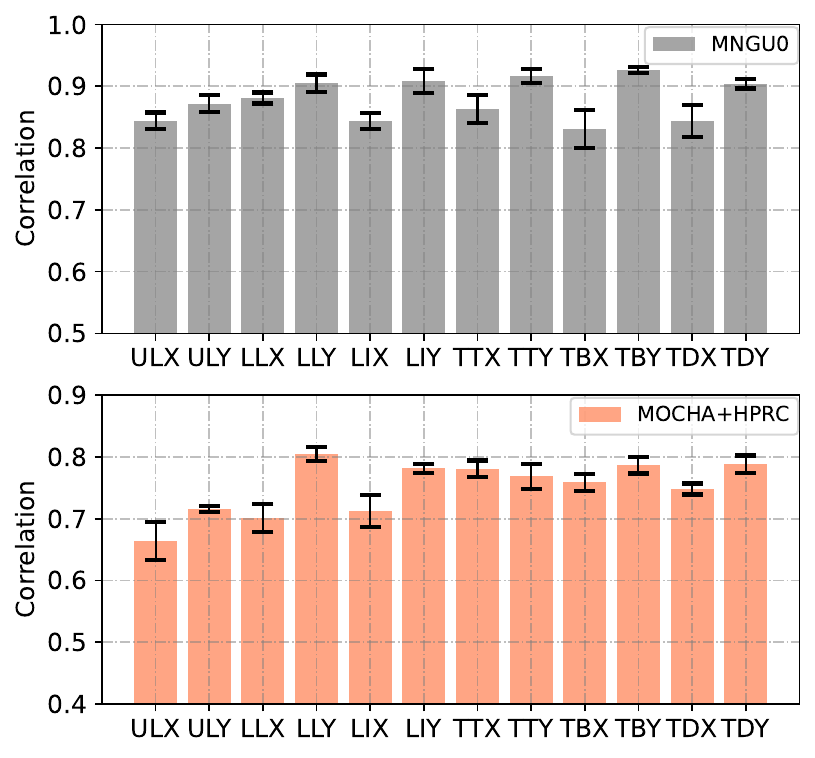}
\vspace{0pt}
\caption{Correlation of individual channels on MNGU (top) and other speakers in MOCHA and HPRC (bottom). }
\label{fig:indiv_channel}
\end{figure}

The performances of AAI on the individual channels are denoted in Fig. \ref{fig:indiv_channel}, measured by correlation (PCC) with 95 \% confidence intervals. The top pannel shows performance on MNGU0 and the bottom channel shows performance after the transformations to other speakers (15 speakers from MOCHA+HPRC). In both cases, the Y-axis traces are generally better predicted than the X-axis traces.

\subsection*{A.4. Evidence of Geometric Isometry Between Individual Articulatory Space}

\begin{figure}[h]
\centering
\includegraphics[width=0.7\linewidth]{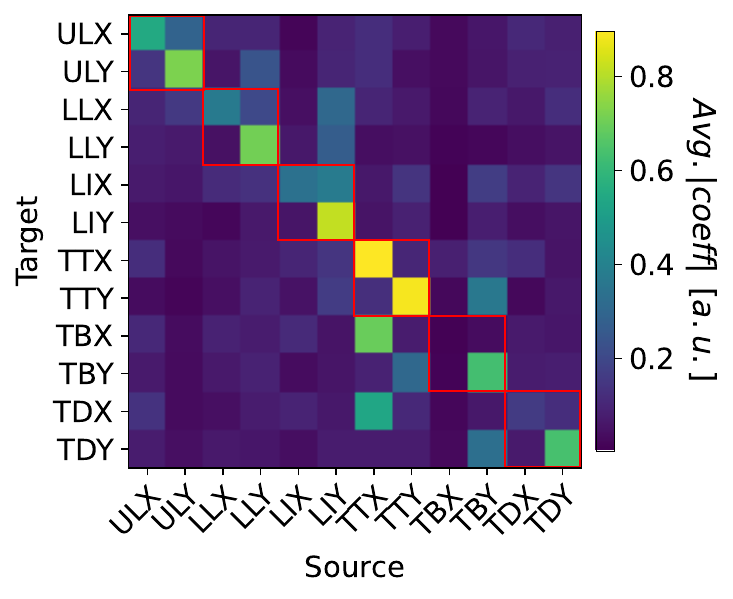}
\vspace{0pt}
\caption{Average absolute coefficients of affine transformation from MNGU0 to other speakers.}
\label{fig:coef_map}
\end{figure}

As shown in \cite{cho2023self}, the weights of the linear transformations across speakers are highly concentrated within the articulators (Fig. \ref{fig:coef_map}; red diagonal boxes). The tongue tip is also affected by the tongue blade and dorsum, which is natural since the tongue is connected and sensor positions are not firm, but still significantly more affected by the tongue tip. 



\subsection*{A.5. Evidence for Excluding Coronal Axes of EMA}

\begin{figure}[h]
\centering
\includegraphics[width=0.7\linewidth]{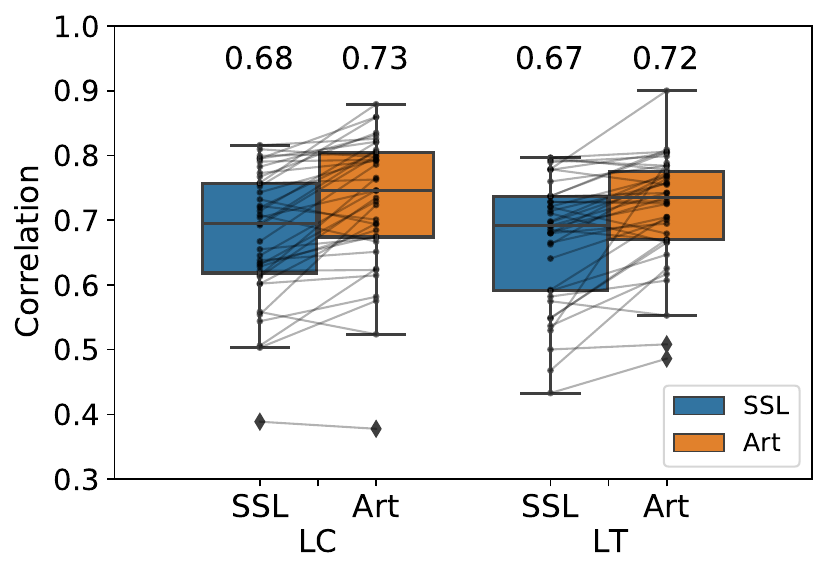}
\vspace{0pt}
\caption{Linear probing on the coronal axis of the left corner of the mouth (LC) and the lateral tongue (LT). The prediction performance by using WavLM features (SSL) and 12 other midsagittal articulatory channels (Art). Each dot is a speaker.}
\label{fig:coronal}
\end{figure}

In our aritculatory encoding, we largely ignore the coronal axes of EMA data. Here, we provide evidence for such exclusion with a probing analysis. The same linear probing is conducted using SSL features (WavLM) and the original 12 midsagittal articulatory channels. We use EMA-MAE \cite{emamae} for this experiment since the dataset includes two coronal channels: the left corner of the mouth (LC) and the lateral tongue (LT). As shown in Fig. \ref{fig:coronal}, both for LC and LT, the linear probing using SSL features is significantly worse than using articulatory features (for LC, SSL features: \(0.68 \pm 0.19\), articulatory features: \(0.73 \pm 0.20\); for LT, SSL features: \(0.67 \pm 0.19\), articulatory features: \(0.72 \pm 0.17\)). This suggests that those coronal features can be predicted with a fair correlation (0.72-0.73) by a linear combination of the existing 12 channels. These results suggest that the coronal axes can be safely ruled out for the proposed coding framework.



\section*{Appendix B -- Model \& Training Detail}

\subsection*{B.1. Pitch Tracker Configuration}

For CREPE, we use a PyTorch implementation called ``torchcrepe" \footnote{Retrieved from https://github.com/maxrmorrison/torchcrepe.}. We use ``full" model and the range of pitch is set as from 50 to 550. The default Viterbi decoding is used. The input waveform is resampled to 16K Hz and z-scored within the utterance. We found that if we set the hop length as 320 to output 50 Hz (the sampling rate of SSL features), sometimes, it makes errors in some female speakers with high pitch levels. To mitigate that we decrease the hop length to 80 which results in 200 Hz outputs, and then, we downsample the output pitch and periodicity by a factor of 4 to match the 50 Hz sampling rate. As a side effect, we observe that the periodicity is sometimes erroneously detected with such hop length modification. Therefore, we threshold the periodicity bigger than 0.4. The periodicity thresholding is only applied for the inference time. 

\subsection*{B.2. Loudness Implementation}

The loudness is estimated as the average of the absolute magnitude of the input waveform for every 20ms bin. This is implemented with a fixed, single-channel convolutional layer with a stride of 320 and kernel size of 320, where each weight in the kernel is set as 1/320. Likewise, the waveform is resampled to 16K Hz and z-scored within utterance.

\subsection*{B.3. Speaker Identity Encoder Architecture}

The CNN outputs of WavLM Large are weighted by periodicity inferred from CREPE and averaged across frames per utterance. Here, we also include the convolutional positional encoding and the projection layer applied for the CNN outputs, in the original WavLM model. That is, we use the features right before entering the Transformer encoder, which has 1024 channels. The FFN is composed of a linear layer that maps 1024 input features to 1024 output features, followed by GELU and dropout with 0.2, and then another linear layer that projects features to 64-channel speaker embedding. This FFN is the only network trained within the vocoder training and other parts are not updated.

\subsection*{B.4. Generator Architecture}

Following the HiFi-GAN architecture, our generator is a convolutional neural network composed of transposed convolutions each followed by a multi-receptive field fusion (MRF) module \cite{kong2020hifi}. Each MRF module outputs the sum of three residual block outputs \cite{he2016resnet}. For our three residual blocks, we use dilations of 1, 3, and 5 and kernel sizes of 3, 7, and 11, respectively. For our transposed convolutions, we use kernel sizes of 10, 8, 4, and 4, and each stride is half the respective kernel size. Channels are halved each layer until the final layer outputs a single channel corresponding to the acoustics.

To condition the generation on speaker identity, a FiLM layer is applied for outputs of each residual convolutional layer in the MRF module \cite{perez2018film}. Every FiLM layer is implemented as a linear layer followed by ReLU and dropout of 0.2, and another linear layer to predict scale and center for each channel of the targeted convolution outputs. These scales and centers are then multiplied and added to the outputs of convolutional layer that the FiLM layer is attached to (feature-wise affine transform).

\subsection*{B.5. Discriminator Architecture}

We utilize two types of discriminators as in the HiFi-GAN model: (1) a multi-period discriminator (MPD), and (2) a multi-scale discriminator (MSD) \cite{kong2020hifi}. The MPD takes as input evenly-spaced input frames, and the MSD average-pools the input. Then, both models feed the processed inputs into convolutional neural networks (CNNs). We follow the same CNN architectures as the HiFi-GAN discriminators \cite{kong2020hifi}. Each MPD CNN is composed of strided convolutional layers each followed by leaky rectified linear unit (ReLU) activation functions. Similarly, each MSD CNN is composed of strided and grouped convolutional layers each followed by leaky ReLUs. Like HiFi-GAN, we use five MPDs with spacings of 2, 3, 5, 7, and 11, and three MSDs that rescale inputs by 1, 2, and 4 times.

\subsection*{B.6. Loss Configuration}

Like HiFi-GAN, our loss function is a weighted sum of the GAN loss, mel-spectrogram loss, and feature matching loss \cite{kong2020hifi}. The weight for three loss terms are 1, 45, and 2, respectively. For the mel-spectrogram loss, we use the following parameters: \{fs: 16000, fft\_size: 1024, hop\_size: 160, win\_length: null, window: "hann", num\_mels: 80, fmin: 0, fmax: 8000\}.

\subsection*{B.7. Training Detail}
We use Adam optimizer with learning rate of \(10^{-4}\) and beta of (0.5, 0.9). The model is updated for 1.5M iterations, and the learning rate is halved by every 8K steps which stays static after 320K steps. For every iteration, a random 320 ms window is sampled from each clip in a batch with a size of 64.

\subsection*{B.8. Multilignual Fine-Tuning Detail}

We fine-tune the model on the multilingual datasets for 500K iterations. The model is initialized with a checkpoint of the model trained only on English for 1M iterations. The training data include 555 hours of English, 1965 hours of German, 1553 hours of Dutch, 161 hours of Portuguese, 247 hours of Italian, 104 hours of Polish, 917 hours of Spanish, 1075 hours of French, 60 hours of Chinese, 22 hours of Japanese, and 11 hours of Korean. The English proportion is the train split of LibriTTS-R.

\bibliographystyle{IEEEtran}
\bibliography{bibliography}

\end{document}